\documentclass{emulateapj}

\usepackage{graphicx} 
\usepackage{color} 
 
\newcommand{\be}{  \begin{eqnarray} }
\newcommand{\ee}{  \end{eqnarray} }

\def\spose#1{\hbox to 0pt{#1\hss}}
\def\lta{\mathrel{\spose{\lower 3pt\hbox{$\mathchar"218$}}
     \raise 2.0pt\hbox{$\mathchar"13C$}}}
\def\gta{\mathrel{\spose{\lower 3pt\hbox{$\mathchar"218$}}
     \raise 2.0pt\hbox{$\mathchar"13E$}}}
\font\syvec=cmbsy10                        
\def\bnabla{\hbox{{\syvec\char114}}}       

\begin{document}
 
\shorttitle{Ultra-Luminous X-ray Event}
\title{Swift J1644+57: An Ultra-Luminous X-ray Event}
\author{Aristotle Socrates}
\affil{Institute for Advanced Study,
Einstein Drive, Princeton, NJ 08540: socrates@ias.edu}

  
\begin{abstract}
The photon spectral energy distribution of the powerful transient Sw
J1644+57 resembles those of the brightest Ultra-Luminous X-ray sources
(ULXs).  The transient nature of Sw J1644+57 is likely the result of a tidal 
disruption of a star by a super-massive black hole.  The stellar disk 
generates accretion power at super-Eddington rates and the observational 
properties of Sw J1644+57 indicate -- in analogy with ULXs -- that the accretion flow
maintains a high level of radiative efficiency with a corresponding super-Eddington
luminosity.  Due to its similarity to ULXs, this powerful 
transient may be thought of as an Ultra-Luminous X-ray event (ULX-E).  
Observational tests for this ULX-E model are proposed as well.  
\end{abstract}

\section{Overview and Observational Facts for Sw J1644+57}

A powerful broad-band source, Sw J1644+57 discovered by 
the {\it Swift} space observatory on March 28, 2011 represents 
genuinely new phenomena.  The combination of its flux, fluence, spectra
and duration are unique and clearly, the excitement that surrounds its
discovery shows the potential of wide-field time domain astronomy.

A compilation of the observed properties of Sw 1644+57 can be 
found in Levan et al. (2011) and Burrows et al. (2011).
The source is broad band, with photon emission from 
radio to the hard X-ray.  By far, the overwhelming majority of
the fluence comes in the form of X-ray emission from $\approx$
0.1-100 keV 
characterized by a flat spectral index $\Gamma\lesssim 2$
that is heavily obscured by absorption at low energies.\footnote{
By definition, when $\nu F_{\nu}=$ a constant, $\Gamma=2$.}
The fact that the {\it Fermi} observatory did not report a 
detection of Sw 1644+57 implies that the high energy 
spectrum is cutoff above {\it Swift}'s Burst Alert Telescope's
energy range.  The source is coincident with the nucleus of 
its presumed host galaxy located at a luminosity 
distance of $1.6$ Gpc.

The temporal properties of the event are bimodal.  There were short-lived
flares, the most powerful and energetic of which lasted $\sim 10^3$ s
with an inferred isotropic X-ray luminosity $L^{\rm flare}_X\sim 3\times
10^{48}$ erg/s.  Over the next $\Delta t\sim 10^6$ s,
the behavior is more steady with $L_X\lesssim 10^{47}$ erg/s and
thus, the ratio of steady to flaring isotropic equivalent
energy release is $\sim 30$.  The inferred energy release during the period
of steady activity $\Delta E_X\sim 10^{53}$ erg, close to the 
1/10 rest mass of a typical star.

From here on, focus is directed primarily towards understanding the physical 
mechanisms behind the production of $\Delta E_X$ worth of X-rays as it is 
responsible for $\sim 95\%$ of the observed energy release.  
The fact that $\Delta E_X$ is close the gravitational binding energy 
of a typical star, at the inner-most circular orbit (ISCO)
of a black hole suggests the following:  (1) the event Sw J1644+57
is powered by black hole accretion (2) the accretion 
is radiatively efficient (3) in conjunction with its transient nature,
the event is likely due to the tidal disruption of 
a $\sim 1M_{\odot}$ star by a super-massive black hole.

As envisioned by Rees (1988), a star becomes tidally disrupted during
close approach to a super-massive black hole, once the orbital separation
becomes comparable to the Roche radius.  For a typical stellar radius
$\sim R_{\odot}$, disruption can only occur for black holes
that are $\lesssim 10^8M_{\odot}$.  Black holes larger than 
$\sim 10^8M_{\odot}$ are unable to produce a tidal disruption 
event -- unless larger stars are considered -- since the star
can be directly swallowed.  

Even if the black hole mass $M_{\bullet}\sim 10^8M_{\odot}$, the
average steady X-ray luminosity $L_X$ exceeds the Eddington Limit,
$L_{_{\rm Edd}}$, by a factor of $\sim 10$.  If the Eddington Limit
cannot be violated, the fluence of Sw J1644+57 cannot be linked to the
source of fuel and energy in a straightforward manner.  The flow may
be out of hydrostatic balance, radiatively inefficient, geometrically
and/or relativistically beamed (Bloom et al. 2011; Burrows et
al. 2011).

The basic physical requirements for black hole accretion to overcome
the Eddington Limit, while maintaining a high level of radiative
efficiency, are outlined by Socrates \& Davis (2006; SD06).  Given
those requirements, they estimate that the spectral energy
distribution of such flows are dominated by a flat $\Gamma\sim 2$
Comptonized X-ray power-law out to $\sim 100$ keV.  With the use of
archival data from {\it XMM}, they show that the brightest
Ultra-Luminous X-ray sources (ULXs) -- the sources that clearly
surpass the Eddington Limit for typical stellar mass black holes --
are, in fact, characterized by flat spectral energy distributions with
$\Gamma\sim 2$.

It therefore seems natural to describe the energy release mechanisms 
of Sw 1644+57 as a scaled-up transient ULX i.e., 
an Ultra-Luminous X-ray event (ULX-E).  In this {\it Letter},
the physical requirements, observational appearance and existing 
evidence of radiatively efficient super-Eddington accretion are given
in \S\ref{s: requirements}.  These principles are applied to Sw J1644+57 
in \S\ref{s: model}.  Observational tests and conclusions are briefly
discussed in \S\ref{s: conclude}

\section{Radiatively Efficient Super-Eddington black hole
accretion: requirements, appearance and existing evidence}
\label{s: requirements}

\subsection{Difficulty in overcoming the Eddington Limit}

For a black hole accretion flow, persistent emission above the 
Eddington Limit is even more prohibitive than say for example, 
a star.  The Eddington Limit results from the 
condition of hydrostatic equilibrium
\be
\frac{\kappa}{c}{\bf F}=\bnabla \phi
\label{e: hydro_stat}
\ee
in the limiting case where the gravitational acceleration
$=\bnabla\phi$ is balanced entirely by the radiation 
force, proportional to the radiative flux ${\bf F}$
and opacity $\kappa$.  The expression above 
can be transformed into the Eddington Limit
\be
\frac{\kappa}{c}L=\frac{\kappa}{c}\int_{\partial V}
d{\bf A}\cdot{\bf F} & = &\int_{\partial V}d{\bf A}\cdot{\bnabla}
\phi=4\pi\,G\int_Vd^3x\rho\nonumber\\
L=L_{_{\rm Edd}}&=&\frac{4\pi\,G\,M_{_{\rm enc}}c}{\kappa}.
\ee
Here, the opacity is taken to be a constant over the Gaussian surface
$\partial V$, which bounds the volume $V$ of the flow.
The mass of the disk is taken to be insignificant in comparison 
to the mass of the black hole so that $M_{\bullet}=M_{_{\rm enc}}$
(see \S II of Abramowicz et al. 1980).  Note that the geometry
of the surface $\partial V$ can be quite arbitrary, implying that
the Eddington Limit is applicable for a disk, rather than just
for an object with spherical symmetry.  Given the mass of 
the black hole $M_{\bullet}$, $L_{_{\rm Edd}}$ is specified by 
the opacity, where $\kappa$ is set equal to the Thomson 
electron-scattering opacity $\kappa_{es}$.

In addition, once the accretion power exceeds Eddington, then 
steady radiatively efficient accretion is also hindered by 
the condition of radiative equilibrium.  For standard 
(thin or slim) $\alpha$-disk accretion (Shakura \& Sunyaev 1973;
Abramowicz et al. 1988), the condition 
is given by 
\be
Q^-_{\rm adv}-\bnabla\cdot{\bf F}=Q^+_{\rm vis}.
\label{e: rad_eq}
\ee
The viscous dissipation rate $Q^+_{\rm vis}$ is balanced by a 
combination of radiative cooling perpendicular to the midplane
and the inward advection of heat, quantified by $Q^-_{\rm adv}$.

SD06 note that the radius at which the heat generated by viscous 
dissipation is advected into the hole, rather than lost by 
radiative diffusion -- mediated by the Thomson opacity $\kappa_{es}$
-- i.e., the trapping radius $R_{tr}$, for values of the accretion
rate above $L_{_{\rm Edd}}/c^2$, is given by
\be
R_{tr}\sim R_g\,\dot{m}
\ee
where $R_g=GM_{\bullet}/c^2$ is the gravitational radius and 
$\dot m$ is the accretion rate in units of $L_{_{\rm Edd}}/c^2$.
For a flow with radiative efficiency $\epsilon$, the ISCO is located
at $R_{in}\sim R_g/\epsilon$.  It follows that as
$L\rightarrow L_{_{\rm Edd}}$, then $R_{tr}\rightarrow R_{in}$.
At the Eddington Limit, the flow marginally maintains a high level 
of radiative efficiency.  Since $R_{tr}\propto\dot{m}$,
the net radiative efficiency decreases in proportion $\dot{m}^{-1}$
and the luminosity is pinned to the Eddington value -- a result of 
the fact that gravitational power is $\propto 1/R$.

\subsection{Physical requirements for surpassing the 
Eddington Limit}

Without addressing any deep theoretical mechanisms (turbulence,
magnetic fields, reconnection, etc.,), it is possible to determine the
necessary physical requirements for a black hole accretion flow to
maintain a high level of radiative efficiency in the event that it
persistently emits above the Eddington value.

In order to circumvent the ``trapping problem,'' SD06 realized that
there must be a form of vertical energy transport that separates
binding energy from the mass of its origin, faster than radiative
diffusion.  An examination of eq. \ref{e: rad_eq} reveals that the
characteristic velocity at which thermal energy can be transported in
the vertical direction by diffusion mediated by electron scattering is
$v_{\rm diff}\sim c/\tau$, where $\tau$ is the Thomson optical depth
of the flow.

By construction, a {\it corona} is a region that possesses relatively 
small amounts of binding energy, but is the recipient of a 
disproportionate amount of it per unit time.  As a result, 
the corona is heated to a temperature above the photospheric
temperature of the optically thick binding energy reservoir.

If the trapping problem establishes the need for a 
corona, the departure from hydrostatic balance inferred from 
eq. \ref{e: hydro_stat}, constrains its optical depth
$\tau_c$.  In the absence of other forces, the photosphere
of the accretion flow cannot be in hydrostatic equilibrium and
a radiation-driven wind is launched from the corona.  
In order to avoid the conversion of photon energy into
outgoing mechanical power of the wind, the optical depth of 
corona $\tau_c\sim 1$ (SD06).  If $\tau_c\gtrsim$ a few,
photons becomes red-shifted as a result of adiabatic losses as they transfer
their energy to matter.  In such an event, the
photon luminosity observed at infinity can be vanishingly small.

\begin{figure}
\epsscale{1.0}
\plotone{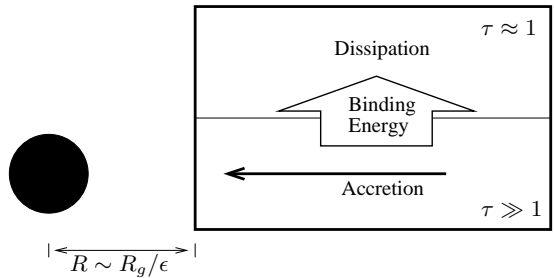}
\caption{Geometry of flow.  Not to scale. }
\label{fig: corona}
\end{figure}

\subsection{Observational Appearance}

Powerful coronae universally appear in the spectral energy 
distribution of sub-Eddington accreting stellar and
super-massive black holes. Though binding 
energy is stored in the ionic disk component, upon dissipation
in the corona gravitational power is transferred to the electrons.
Comptonization by seed photons results in a power law distribution 
of X-ray photons, indicative of a second order Fermi process
(Kompane'ets 1956; Sunyaev \& Titarchuk 1980).  

The spectral shape in the limit of un-saturated Comptonization by a
thermal plasma is primarily determined by the Compton {\it
y-}parameter given by $y=(4k_BT_e/m_ec^2)\tau_{c}$ in the limit that
the corona is optically thin.  A potential source of seed photons is
the transitional region between the corona and the dense disk (Ross \&
Fabian 1993; Ross et al.  1999; Ballantyne et al. 2001; Nayakshin et
al. 2000; Nayakshin \& Kallman 2001).  The cold dense disk is not a
perfect reflector and down-ward Comptonized X-rays serve as a source
of heat due to non-zero albedo.  Above photon energies of 30 keV,
Compton down-scattering dominates the albedo while at lower energies,
photo-electric absorption mainly contributes to the albedo.  These
calculations show that for high values of ionizing flux, which is
$\propto L_c$, fluorescence and absorption features in the X-ray
spectrum persist, though at a highly suppressed level.  The outgoing
spectrum is roughly represented by a flat power law that extends to
the cut-off energy given by $k_BT_e$.  For AGN, the power law spans
from tens of eV all the way to several hundreds of keV.\footnote{The
author thanks J. Goodman and J. Gunn for pointing out that the
observed spectrum, in conjunction with the criteria that $\tau_c\sim
1$, implies that thermal disk photons cannot be responsible for the
seed photons for bright ULXs -- in contradiction to the assertions of
SD06.  In the Appendix it is further shown that thermal disk emission
cannot escape into the corona as it is trapped inward to due to
advection.}

\subsection{Existing Evidence:  Bright ULXs}

The brightest ULXs with inferred bolometric
luminosities $\sim 10L_{_{\rm Edd}}$ for a stellar
mass black hole are dominated by hard power-law X-rays.
Such behavior is in line with the theoretical 
expectations outlined above.

In sub-Eddington black hole accretion flows that power
black hole binaries (BHBs) and AGN, the fraction of energy 
dissipated in the corona {\it f} is roughly given by 
the ratio of Comptonized to bolometric power.  Table
\ref{t: states} shows that for BHBs, the ratio $L_c/L$
is correlated with $L/L_{_{\rm Edd}}$.  For the ``ULX state''
(SD06), the black hole mass is not measured and is taken 
to have a characteristic value of $10\,M_{\odot}$.  

Furthermore, Grimm et al. (2003) construct the combined luminosity 
function of local star-forming galaxies.  When they restrict their
study to high mass X-ray binaries (HMXBs; companion mass $> 5\,M_{\odot}$),
they find that the luminosity function is well-described by a single 
power-law up to (un-extrapolated) luminosities of $3\times 10^{40}$ erg/s
i.e., 30 times the Eddington Limit for a 10$M_{\odot}$ black hole
(see their figure 5).  If these spectra were modeled and extrapolated 
to higher energies (100 keV), then their luminosity would extend 
to $\sim 10^{41}$ erg/s.  Both properties of their luminosity
function simultaneously dictate that super-Eddington accretion is
at work and that it is radiatively efficient.

If intermediate mass black holes (IMBHs) are responsible for the
population of extra-galactic HMXBs above the characteristic Eddington
Limit, then one would expect a break in the luminosity function.
It is unreasonable to expect that IMBHs are members of binary systems
in the same number as stellar mass black holes.  If the high end
of the HMXB luminosity function signals super-Eddington luminosities
from radiatively {\it in-efficient} accretion, then luminosity 
function would exhibit a break above $L_{_{\rm Edd}}$.  Similarly,
a beamed-jet explanation for the high end of the HMXB luminosity function would
produce a break in the luminosity function.  The fact that the luminosity
function is a power-law is most easily interpreted by realizing that 
the central sources of HMXBs are wind-fed and the rate at which 
high mass stars donate mass to their compact companions is set by 
stellar evolution.

The fact that the spectral energy distributions of the bright ULXs
possess the same properties -- from basic theoretical arguments and
observational expectations -- as that of super-Eddington radiatively
efficient accretion flows is suggestive.  In addition, the luminosity
function of the extra-galactic HMXBs reinforces the notion that bright
ULXs represent a mode of black hole accretion that is both
super-Eddington and radiatively efficient.

\begin{deluxetable}{ccc}
\tabletypesize{\scriptsize}
\tablecaption{Spectral Classification of Luminous Black Hole 
Binaries Based on Coronal Behavior
\label{t: states}}
\tablewidth{0pt}
\tablehead{\colhead{Spectral State} &\colhead{$L/L_{_{\rm Edd}}$} & \colhead{$L_c/(L-L_c)$}}
\startdata
High/Soft & 0.1 & 0.1
\\
Very High & 1.0 & 1.0
\\
ULX & 10.0 & 10.0
\enddata 
\tablenotetext{a}{The values listed above are approximate.  Fractional 
coronal power
is correlated with bolometric luminosity.}
\end{deluxetable}

\section{model for Sw J1644+57: An Ultra-Luminous X-ray Event}
\label{s: model}

In its observational appearance as seen through its spectral 
energy distribution, Sw 1644+57 can be thought of as a scaled-up
bright ULX.  Again, the bright ULXs represent a class of 
stellar mass sources objects that persistently emit well
above the Eddington Limit with high radiative efficiency.
Therefore, it seems natural to construct an accretion model 
of Sw J1644+57 that resembles those of bright ULX,
multiplied by a factor of $\sim 10^6-10^7$ in mass.

Rather than steady feeding from a companion star,
the source of fuel results from the tidal disruption 
of an ordinary star (Rees 1988).  There has been recent theoretical
progress on this topic (Rosswog \& Ramirez-Ruiz 2009), yet the
resulting spatial distribution of the debris is remains
uncertain.  Assume that $\sim 1 M_{\odot}$ of stellar gas
is placed in a circular orbit at radius $R_0$ that is some
multiple $\beta$ of the ISCO located at $R_g/\epsilon$.  
Within a dynamical time, the disk becomes viscous and 
-- for lack of a better choice -- the viscosity can 
be approximated by the usual $\nu\sim \alpha c_s H$
(Shakura \& Sunyaev 1973). The viscous time 
$t_{\nu}$ determines the characteristic duration of the event.  Since
the radiation cooling time is long in comparison to the viscous 
time, the flow is likely to be geometrically thick i.e., 
the vertical scale height $H\sim R$ with sound speed $c^2_s\sim
GM_{\bullet}/R$.  In terms of these parameters,the viscous
time is given by
\be
t_{\nu}& = & R^2_0/\nu\sim \frac{\beta^2R^2_g}{\epsilon^2\alpha\,c_s\,H}
\sim \frac{\beta^{3/2}\,R_g}{\epsilon^{3/2}\alpha\,c}\nonumber\\
t_{\nu}& \sim & 3\times 10^6\frac{\beta^{3/2}_{10}\,M_{\bullet ,8} }{
\epsilon^{3/2}_{0.1}\alpha_{0.1}}\,{\rm s}.
\ee
Reasonable changes in $\epsilon$ and $\beta$ lead to comparable values
of $t_{\nu}$ for a variation of $\sim 10$ in $M_{\bullet}$.  Reported
variability timescales of $\sim 100$ s indicate that 
$M_{\bullet}\sim 10^7M_{\odot}$.  The size of the host galaxy further 
supports the ``small'' black hole hypothesis.

If $\sim 1\,M_{\odot}$ worth of material is swallowed with radiative
efficiency $\epsilon\sim 0.2$, then the event can support a
luminosity $L\sim 10^{47}$ erg/s for $3\times 10^6$ s.
If $M_{\bullet}\sim 10^7M_{\odot}$, then $L/L_{_{\rm Edd}}$
is consistent with the high end of Eddington ratios for extra-galactic
HMXBs.

Since $L>L_{_{\rm Edd}}$, hydrostatic balance is broken in the absence
of other forces (e.g., magnetic fields anchored deep in the cold dense
disk SD06).  A Compton radiation-driven wind develops and even for a
small black hole with $M_{\bullet}\sim 10^7M_{\odot}$, the total
compactness $\ell\equiv\epsilon L/L_{_{\rm Edd}}$ is only $\sim 10$.
Under these conditions, a Compton-driven wind is inefficient at
converting the input photon luminosity into bulk wind power (Madau \&
Thompson 2000).  On energetic grounds, a heavily Comptonized medium is
unable to efficiently launch a powerful wind since, by construction,
the electrons are donating energy to the photons as they attempt to they
gain momentum from them (Phinney 1982).

Along with the low coronal optical depth $\tau_c$, which implies
that adiabatic losses are not catastrophic, it seems reasonable 
that a $\sim 10^7\,M_{\odot}$ black hole can accrete a stellar
mass in $\sim 10^6$ s with high radiative efficiency as long
as a fraction $f$ of the binding energy dissipated in the corona
is $\approx 1$.  The observed shape of the photon spectrum 
resembles those of the brightest ULXs i.e., a flat 
power-law in X-rays.

\section{Questions, Predictions and Conclusions}
\label{s: conclude}

The central energy source of Sw J1644+57 is heavily obscured.  It is
almost a certainty that, due to energy conservation, all of
dust-obscured photon power will be re-emitted in the IR wavebands at
some later time.  If the source is an isotropic emitter, such as the
ULX-E model presented here, the total energy emitted in the IR is
likely to exceed that of a beamed jet model.  If a sizeable fraction
of $10^{53}$ ergs is absorbed by dust, then the reprocessed IR
emission will easily out-shine the host as long as the thermal
relaxation time of the dusty medium does not exceed several decades.

As stellar debris drains into the black hole, the rate of accretion
drops, as does $\tau$ of the dense disk and the bulk of the 
flow is eventually able to radiate efficiently.  Sub-Eddington
AGN and BHBs possess spectra that show strong thermal emission.
Once the accretion power drops below the Eddington Limit, 
the characteristic energy of the disk photons $\sim 50$ eV for 
$M_{\bullet}\sim 10^7M_{\odot}$. There may be enough optical-UV power
in the thermal component at later times so that the central 
source out-shines the host galaxy.

Epochs of intense and highly variable flaring X-ray power as well as
radio emission (Giannios \& Metzger 2011), are most likely due to the
production of a jet at early times.  {\it Swift}'s wide-field BAT
instrument was triggered by these bright flaring events, rather than
the relatively dim persistent radiative emission responsible for the
bulk of the energy release $\Delta E_X$.  If the jet opening-angle is sufficiently
small, then the rate of tidal disruption events are in line with
theoretical expectations (Bloom et al. 2011; Burrows et al. 2011; see
Magorrian \& Tremaine 1999) .

Even for sub-Eddington accretion flows, the nature of the transport
mechanism that energizes the corona has not been identified for
thirty-five years (e.g., Shapiro et al. 1976).  Possibilities include
magnetic buoyancy (Galeev et al. 1979), wave excitation and turbulence
(Socrates et al. 2004; Thompson 2006).  The high compactness of black
hole and neutron star accretion flows in comparison to other
astrophysical systems may explain why their inferred value of $f$ is
so high (Goodman \& Uzdensky; Socrates 2010).  Nevertheless, the
distance between a robust and testable theory that quantifies the
fraction {\it f} of binding energy dissipated in the corona and our
current understanding seems far.  The possibility that Sw J1644+57 is
powered by coronal energization reinforces the general desire to
identify the physical mechanisms that power Comptonized radiation of
relativistic accretion flows.

\acknowledgements{It is a pleasure to thank J. Miradla-Escude,
E. Ramirez-Ruiz and R. Sunyaev for helpful conversations as well as
D. Giannios and B. Metzger for comments on the manuscript.  Support
form a John Bahcall Fellowship awarded by the Institute for Advanced
Study, Princeton is acknowledged. }

\appendix

\section{The dense disk must be advection dominated}

Let $t_{\rm adv}\equiv R/v_R$ and $t_{\rm diff}\equiv \tau H/c$ where
\be
v_{R}=\frac{\dot{M}}{4\pi R\,H\,\rho}=\frac{\dot{M}\kappa}{4\pi\,R\,\tau}
\ee
and $\dot{M}$ is the accretion rate.  Compute the ratio
\be
\frac{t_{\rm adv}}{t_{\rm diff}}=\frac{4\pi\,R^2\,\tau}{\dot{M}\kappa}\frac{c}{\tau\,H}
=\frac{4\pi\,R\,\epsilon\,c^2 }{L\kappa}.
\ee
Most of the energy is liberated near the ISCO, located at $R\sim R_g/\epsilon=
GM_{\bullet}/\epsilon c^2$ and at that location
\be
\frac{t_{\rm adv}}{t_{\rm diff}}=\frac{R}{R_g}\frac{\epsilon}{L/L_{_{\rm Edd}}}
\sim \frac{L_{_{\rm Edd}}}{L}.
\ee
The cool dense disk it too optically thick to cool when $L\gg L_{_{\rm Edd}}$
even if the gravitational power it generates is dissipated exterior to it, in 
the adjacent corona.  It follows that due to the long diffusion time, the 
disk cannot serve as a source of seed photons for the super-Eddington
corona, in contradiction to the claim of SD06.  In other words, the fraction
of gravitational power dissipated in dense the disk $=(1-f)$ cannot escape via 
radiative diffusion.  Once the accretion power becomes super-Eddington, the optically
thick thermal emission component generated near the ISCO disappears and is 
swallowed into the black hole.  Only the fraction $f$ of the binding energy deposited in 
the corona is available for the production of escaping radiation.  In order for
the flow to maintain a high level of radiative efficiency above the Eddington
Limit, $f\approx 1$.

\end{document}